\documentclass[twocolumn,showpacs,prX]{revtex4-1}

\usepackage{ctable}
\usepackage{rotating}
\usepackage{epstopdf}
\usepackage{subfigure}
\usepackage{enumitem}
\usepackage{verbatim}
\usepackage{eqparbox}

\graphicspath{{figures/}}

\newcommand{\specialcell}[2][c]{\footnotesize \begin{tabular}[#1]{@{}l@{}}#2\end{tabular}}
  
\begin{document}

\title{Complete topology of cells, grains, and bubbles \\in three-dimensional microstructures}

\author{Emanual A. Lazar$^1$, Jeremy K. Mason$^2$, Robert D. MacPherson$^1$, David J. Srolovitz$^3$}
\affiliation
{$^1$School of Mathematics, Institute for Advanced Study, Princeton, New Jersey 08540, USA\\ 
 $^2$Lawrence Livermore National Laboratory, Livermore, California 94550, USA\\
 $^3$Institute of High Performance Computing, 1 Fusionopolis Way, 16-16 Connexis, Singapore 138632} 
\date{\today}

\begin{abstract}  
We introduce a general, efficient method to completely describe the topology of individual grains, bubbles, and cells in three-dimensional polycrystals, foams, and other multicellular microstructures.  This approach is applied to a pair of three-dimensional microstructures that are often regarded as close analogues in the literature: one resulting from normal grain growth (mean curvature flow) and another resulting from a random Poisson--Voronoi tessellation of space.  Grain growth strongly favors particular grain topologies, compared with the Poisson--Voronoi model.  Moreover, the frequencies of highly symmetric grains are orders of magnitude higher in the the grain growth microstructure than they are in the Poisson--Voronoi one.  Grain topology statistics provide a strong, robust differentiator of different cellular microstructures and provide hints to the processes that drive different classes of microstructure evolution.
\end{abstract}


\maketitle

Characterizing the microstructure of materials has occupied an important place in much theoretical, experimental, and computational work over the last fifty years.  Such microstructures include the cellular structure of foams, polycrystalline materials and biological systems.  Different cellular structures share many features in common, yet even rudimentary analysis shows that structures resulting from different formation or evolution processes can be startlingly different.    For example, Poisson--Voronoi tessellation of space yields a microstructure that is akin to those produced by crystallization or recrystallization \cite{1998gough}, while structures that evolve through curvature flow describe normal grain growth structures \cite{2004Humphreys}.  Despite the differences in the resulting structures, the former is often assumed to accurately represent experimental systems, despite the fact that the latter may be more suitable. This is important because such structural differences can lead to markedly different physical properties.

\begin{figure}[h]
\begin{center}    
\begin{tabular}{cc}      
{\resizebox{0.23\columnwidth}{!}{\includegraphics{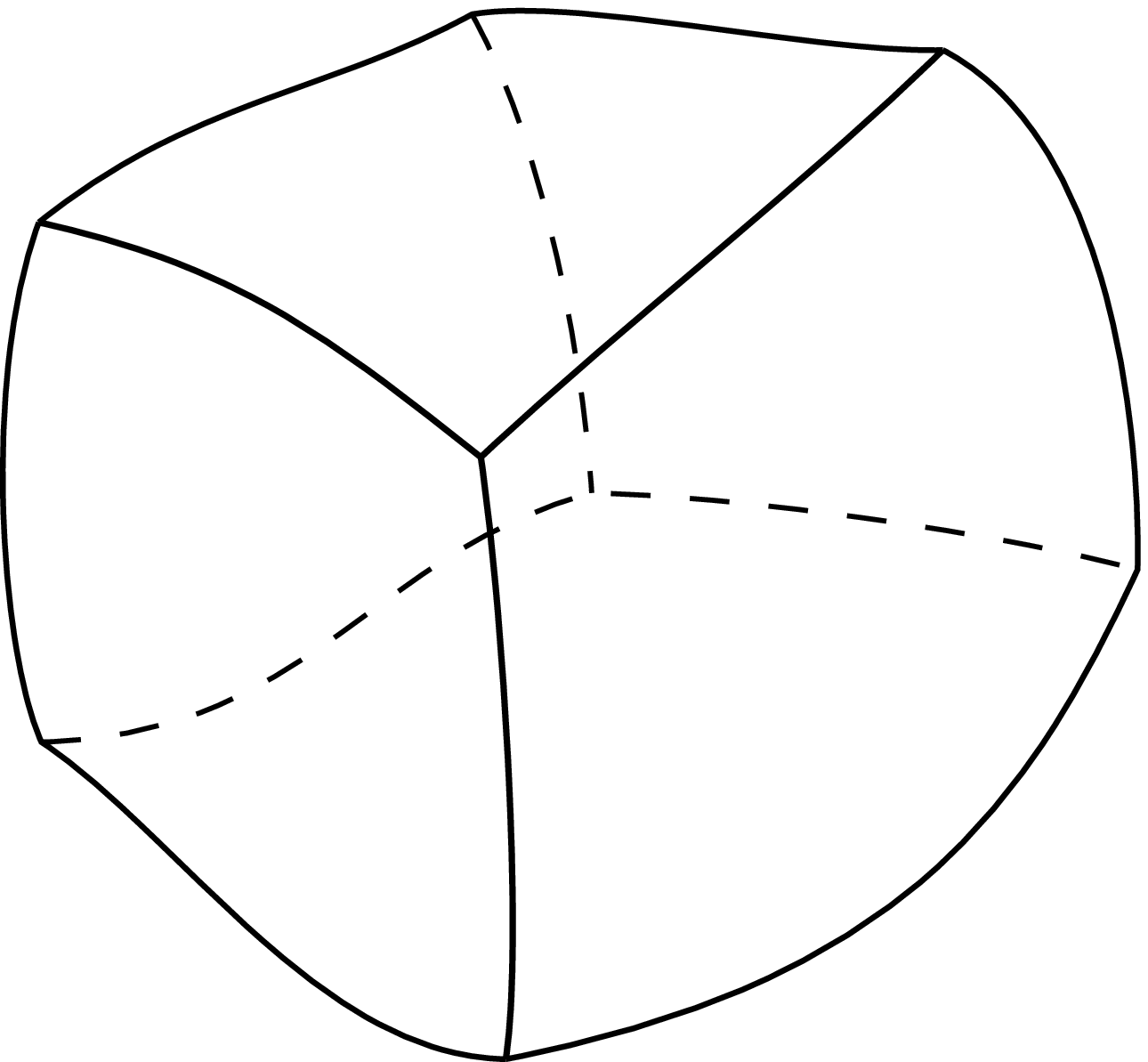}}} &\quad       
{\resizebox{0.20\columnwidth}{!}{\includegraphics{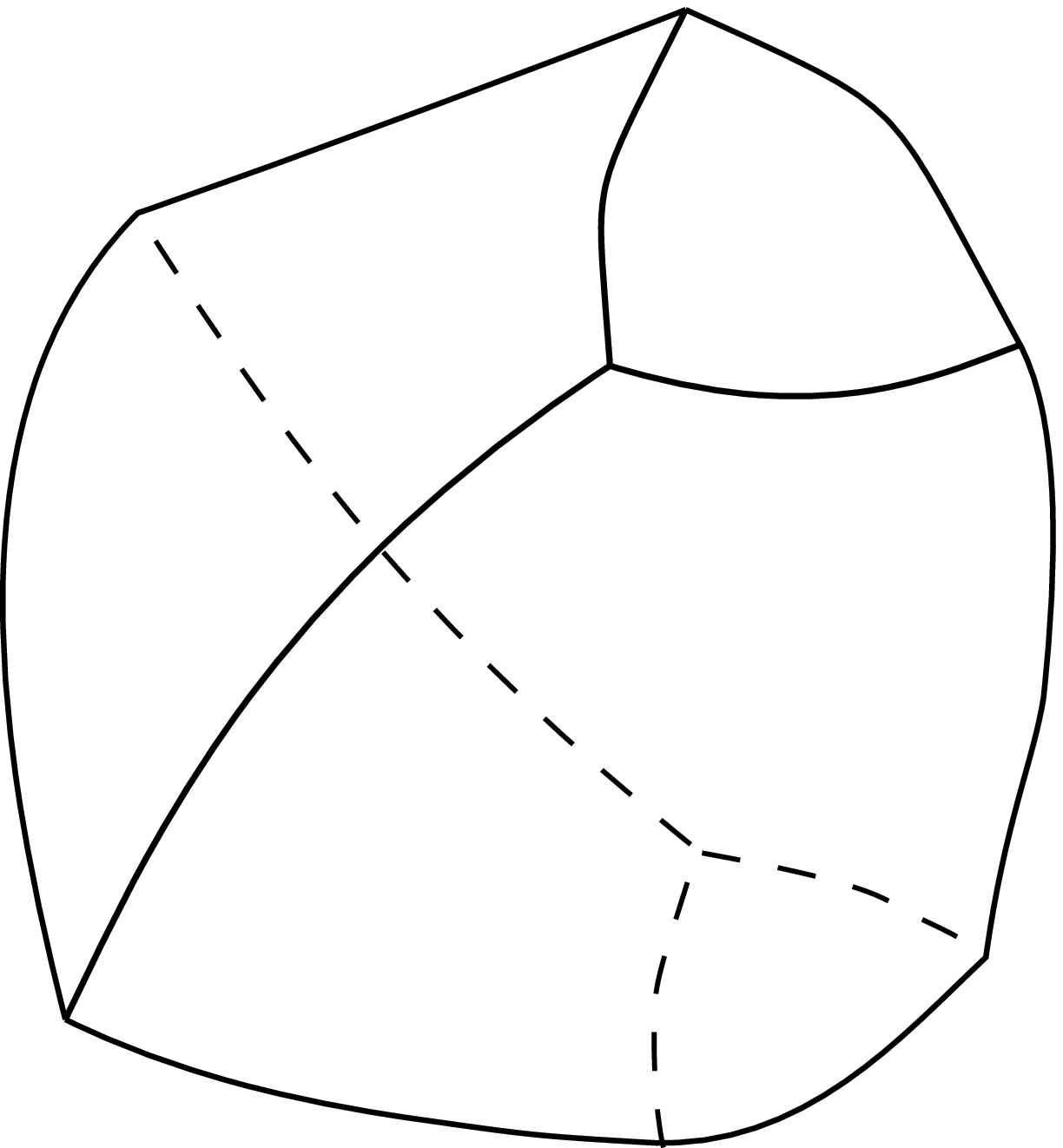}}}
\end{tabular}    
\caption{Two topologically distinct 6-faced grains.} 
\label{sixfaceshapes}
\end{center}
\vspace{-5mm}
\end{figure}

Describing grains of such microstructures involves measuring not only their geometric features such as mean cell size and aspect ratio, but also their topological features.  Historically, the topology of an individual grain has been commonly described using only its number of faces (e.g., see \cite{2007rios, 1974rhines}); this is motivated both by the fact that this is a relatively straightforward measurement and because of its analogy to key features of the rigorous theory of two-dimensional grain growth \cite{1952vonneumann, 1956mullins}.  The simplicity of this characterization allows for easy gathering and succinct presentation of data and has been widely applied experimentally and in theory and simulations. While the number of faces of a grain is a basic topological descriptor, it is clearly incomplete; consider the two topologically distinct 6-faced grains in Fig.~\ref{sixfaceshapes}. In this Letter, we present a new method of completely describing the topology of grains or cells within three-dimensional microstructures and apply it to show several stark differences between the Poisson--Voronoi and normal grain growth microstructures.

A more detailed topological description was introduced by Matzke \cite{1946matzke} for bubbles in soap foams.  Matzke characterized a large population of  bubbles by recording the total number of faces and the number of edges of each face in each bubble.  This method distinguishes between the two grains in Fig.~\ref{sixfaceshapes}---the first  contains six quadrilateral faces; the second two triangular, two quadrilateral, and two pentagonal faces.  We associate a vector of non-negative integers with each grain: the $i^{\text{th}}$ entry counts the number of $i$-sided faces.  Following \cite{1969barnette}, we call this the {\bf $p$-vector} of a grain.  The grains illustrated in Fig.~\ref{sixfaceshapes} have distinct $p$-vectors $(0022200...)$ and $(0006000...)$.  This facilitates a more detailed characterization of the topology of a grain than does recording only its number of faces; e.g., it allows the determination of the fraction of 12-faced grains that are pentagonal dodecahedra.  

Although powerful, this approach has not been widely applied.  Historically, obtaining and analyzing large grain growth or bubble microstructures (by experiment or simulation) has been quite difficult.  Many recent large grain growth simulations use methods that do not directly lend themselves to topological analysis---phase field \cite{2006kim}, Monte Carlo \cite{2006zollner} and diffusion-based \cite{2011elsey} simulations employ implicit descriptions of grain shape, complicating accurate topological analysis. On the other hand, front-tracking, three-dimensional grain growth simulations \cite{2000wakai, 2011lazar} produce large microstructures from which grain topology may be readily extracted \cite{2012lazar}.  The statistics reported below include contributions from 25 steady-state normal grain growth microstructures \cite{2012lazar}, each containing about 10,782 grains (i.e., 269,555 grains in total).  A Poisson--Voronoi microstructure, generated by a Voronoi tessellation of 269,555 Poisson-distributed points in the periodic unit cube, is used as a comparative microstructure.  

\begin{table}
\centering
\makebox[0pt][c]{\parbox{0.8\columnwidth}{
\begin{minipage}[b]{0.45\hsize}\centering
\begin{tabular}{ | c | c | c |}
\hline
\multicolumn{3}{| c | }{{\bf Poisson--Voronoi }}\\ 
\hline
$F$ & {\bf $p$-vector} & $f(\%)$ \\ 
\hline
12 & (001343100...) & 0.39 \\
11 & (001342100...) & 0.33 \\
13 & (001433200...) & 0.30 \\
13 & (002333110...) & 0.29 \\
9 & (001332000...) & 0.28 \\
13 & (001344100...) & 0.28 \\
13 & (001352200...) & 0.28 \\
11 & (001423100...) & 0.28 \\
\hline        
\end{tabular}
\end{minipage}
\hfill
\begin{minipage}[b]{0.45\hsize}\centering
\begin{tabular}{ | c | c | c |}
\hline
\multicolumn{3}{| c | }{{\bf Grain-growth}}\\ \hline
$F$ & {\bf $p$-vector} & $f(\%)$ \\ \hline
8&(000440000...) & 2.83 \\
10&(000442000...) & 2.38 \\
9&(000360000...) & 1.86 \\
11&(000443000...) & 1.86 \\
9&(000441000...) & 1.63 \\
7&(000520000...) & 1.48 \\
12&(000363000...) & 1.45 \\
10&(000361000...) & 1.43 \\
\hline
\end{tabular}
\end{minipage}
}}
\caption{Lists of the eight most common $p$-vectors, their number of faces $F$, and their frequencies \emph{f} in the Voronoi and grain growth microstructures.  The relative errors in the $p$-vector frequencies are less than 4\% for the Poisson--Voronoi microstructure and less than 2\% for the grain growth one.}
\label{pvectortable}
\end{table}

Table \ref{pvectortable} enumerates the most common $p$-vectors in the Poisson--Voronoi and grain growth microstructures.  Two differences between the microstructures are readily apparent.  All of the frequently occurring $p$-vectors in the Poisson--Voronoi microstructure contain at least one triangular face.  By contrast, not one of the frequently occurring $p$-vectors in the grain growth structures contains a triangular face.  Also, almost all frequently occurring $p$-vectors in the Poisson--Voronoi structure contain at least one heptagonal face, while the frequently occurring $p$-vectors in the grain growth structures do not.

One might attribute this discrepancy to the higher frequency of triangular faces in the Poisson--Voronoi microstructure than in the grain growth microstructure, since $13.5\%$ of all faces are triangular in the former and only $4.3\%$ are triangular in the latter \cite{2012lazar}.  However, this fails to account for differences with respect to heptagonal faces.  Whereas $11.6\%$ of faces are heptagonal for the Poisson--Voronoi microstructure, the corresponding frequency is $8.4\%$ for the grain growth case.  
Although these frequencies differ by less that  $50\%$, almost all of the most frequent $p$-vectors in the Poisson--Voronoi structure have some heptagonal faces, whereas none of those in the grain growth structure have.  This large difference cannot be accounted for by the difference in the frequencies of heptagonal faces alone.

A second and perhaps more striking difference between the structures is the manner in which $p$-vectors are distributed.  In the grain growth microstructure, the eight most common $p$-vectors account for almost $15\%$ of all grains, while they account for less than $2.5\%$ in the Poisson--Voronoi case.  Since the process of normal grain growth drives the reduction in grain boundary area per unit volume, this favors more equiaxed grains.  Presumably, this is achieved more readily with certain combinations of polygons on the surfaces of grains than others, leading to the observed selectivity of the grain growth process.

\begin{figure}[t]
\begin{center}
\begin{tabular}{cc}
\subfigure{\label{2222a}{\includegraphics[width=0.205\columnwidth]{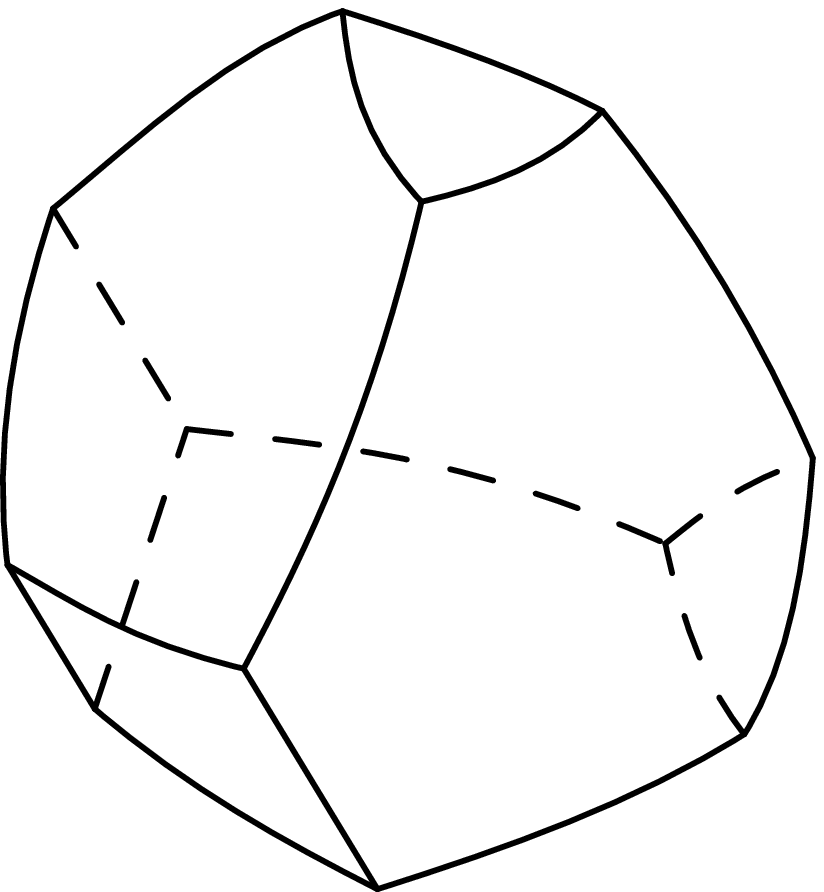}}} &\quad
\subfigure{\label{2222b}{\includegraphics[width=0.23\columnwidth]{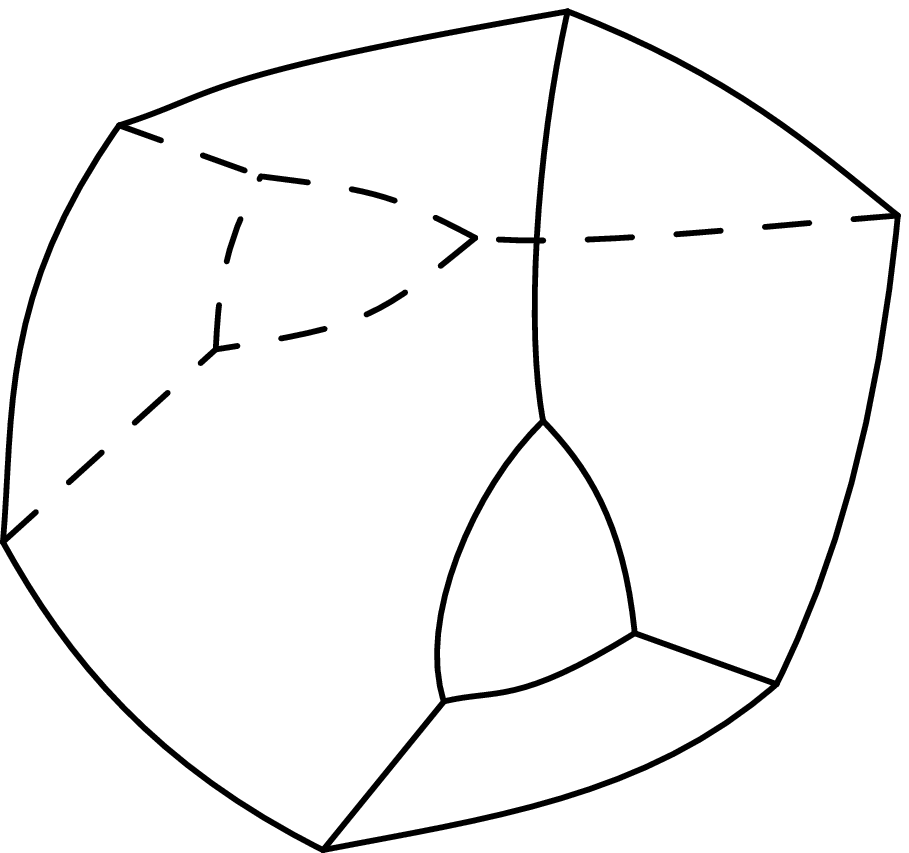}}}
\end{tabular}
\caption{Two topologically distinct grains that share $p$-vector $(00222200...)$.  In the first, two triangular faces are connected by an edge, whilst in the second they are not. 
}
\label{2222}
\end{center}
\vspace{-5mm}
\end{figure}

Although a $p$-vector offers a more refined description of a grain than a mere count of its faces, it too is incomplete.  Consider that a fixed set of polygonal tiles can be arranged on the boundary of a grain in multiple topologically-distinct ways. Figure~\ref{2222} illustrates two such distinct grains which share a  $p$-vector.  

A complete characterization of three-dimensional grain topology can be built on Weinberg's work \cite{1966weinberg1, 1966weinberg2} on determining if two triply-connected planar graphs are isomorphic.  He introduced an encoding of the topological structure of such graphs into a vector of integers and showed that the two graphs are isomorphic if and only if these vectors  are identical.  We employ this approach to encode the topology of each grain and use these to catalogue distinct topological types and their frequencies in various microstructures. 

\begin{figure}[b]
\centering
\includegraphics[width=0.22\textwidth]{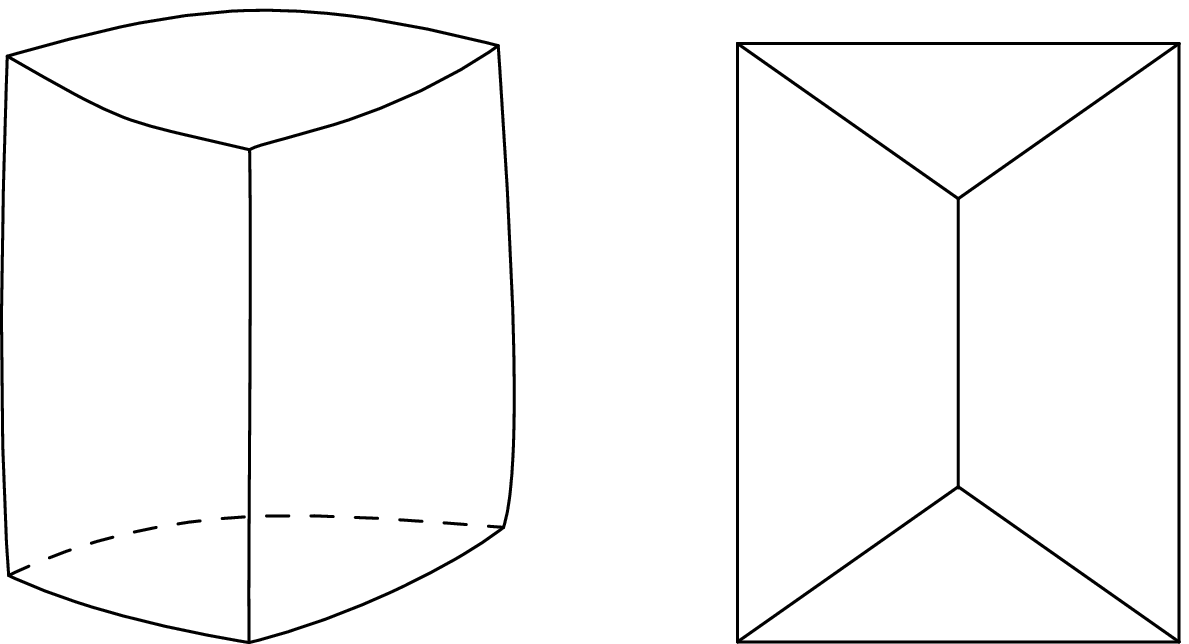}
\caption{A typical grain and a corresponding Schlegel diagram (the embedding of the vertex-edge graph in the plane).}
\label{typical_grain}
\end{figure}

The first step is to reduce a three-dimensional grain topology to a vertex-edge planar graph---a Schlegel diagram \cite{1995Ziegler}---as shown in Fig.~\ref{typical_grain}.  The Schlegel diagram can be constructed by projecting the polyhedron onto one of its faces (the vertices which do not belong to that face lie inside the face onto which the polyhedron is projected). This allows us to use ``right turn'' and ``left turn'' unambiguously when ``traveling'' along a path in the graph. An initial vertex is chosen and assigned the label 1; that label is appended to an initially empty vector.  Next, one of that vertex's three adjacent edges is chosen and travel begins along that edge.  These rules are then followed:
\begin{enumerate}[leftmargin=*]
\item When an unlabeled vertex is reached, label it with the next largest unused integer, append that label to the vector, and then ``turn right'' and continue traveling in the graph.
\item When a previously-labeled vertex is reached, append its label to the vector and then:
\begin{enumerate}[leftmargin=*]
\item If this vertex is reached by traveling along an edge that has not been traversed, return to the previous vertex along the same edge but in the opposite direction.
\item If this vertex is reached by traveling along an edge that has been traversed (in the opposite direction), ``turn right'' and continue traveling; if that right-turn edge has previously been traversed in that direction, ``turn left'' and continue traveling; if that left-turn edge has also been traversed, stop.  
\end{enumerate}
\end{enumerate}

\begin{figure}
\centering
\includegraphics[width=1.0\columnwidth]{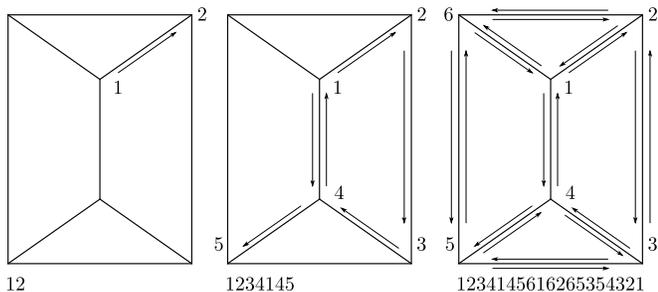} 
\caption{Vertices are labeled as they are initially encountered while traversing the graph following rules (1) and (2); the vector lists all vertices in the order in which they have been visited.}
\label{gull}
\end{figure}

Fig.~\ref{gull} illustrates the process of constructing a vector for the graph of a typical grain.  An initial vertex and edge are chosen, and a path through the graph is followed according to rules (1) and (2); vertices are labeled in the order they are visited.  The vector for this particular path in the grain is 1234145616265354321.  

When we stop, each edge has been traversed exactly once in each direction; each vertex has been visited exactly three times (starting/ending vertices are visited four times).  The path through the graph is recorded as a vector of vertex labels.  This vector depends only on the topological structure of the grain and on the choice of initial vertex and edge, but not on any prior labeling of the vertices.

Other vectors for a grain are produced by repeating this procedure for other initial vertices and initial edges, and likewise for the mirror image of the grain.  This procedure yields $4E$ vectors for each graph with $E$ edges; each vector is a list of $2E+1$ integers.   Because the grain topology can be reconstructed from any  of the vectors, we need only record one vector per grain.  We call the lexicographically (numerical/alphabetical order) first vector the {\bf Weinberg vector} of the grain.

Although $4E$ vectors are constructed for each grain, vertex-edge graphs containing mirror and rotational symmetries will have fewer unique ones. The order $S$ of the symmetry group associated with a grain is the number of total vectors divided by the number of unique vectors.  The grain in Figs.~\ref{typical_grain} and \ref{gull} contains $E=9$ edges and only 3 unique vectors; hence, the order of its symmetry group is $S = 4E / 3 = 12$.  

\setlength{\tabcolsep}{6pt}
\begin{figure*}[ht]
\centering
\begin{tabular}{|c|c|c|c|c|c|c|c|}
\hline
\multicolumn{8}{| c | }{{\bf Poisson--Voronoi }}\\ 
\hline
\specialcell[b]{P1, $f$=0.28\%\\\includegraphics[scale=0.24]{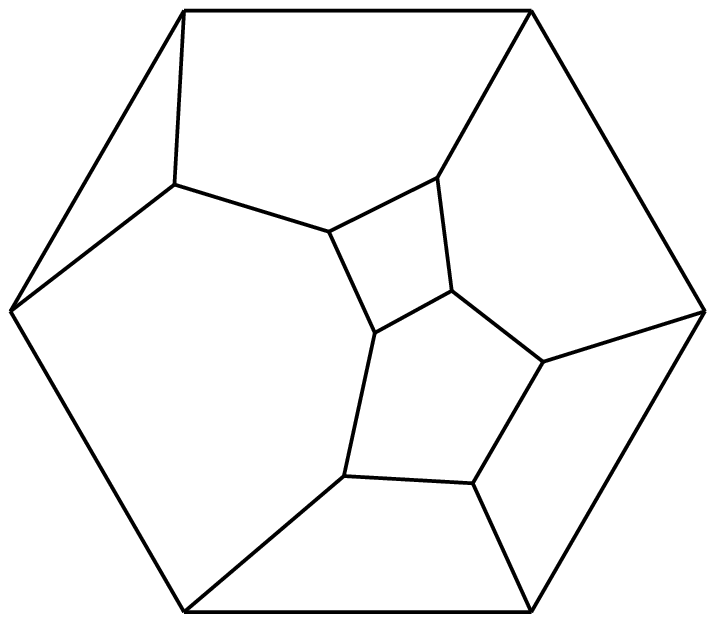}\\(00133200...)\\\hspace*{1.5mm}$F$=9, $S$=1 } & 
\specialcell[b]{P2, $f$=0.17\%\\\includegraphics[scale=0.24]{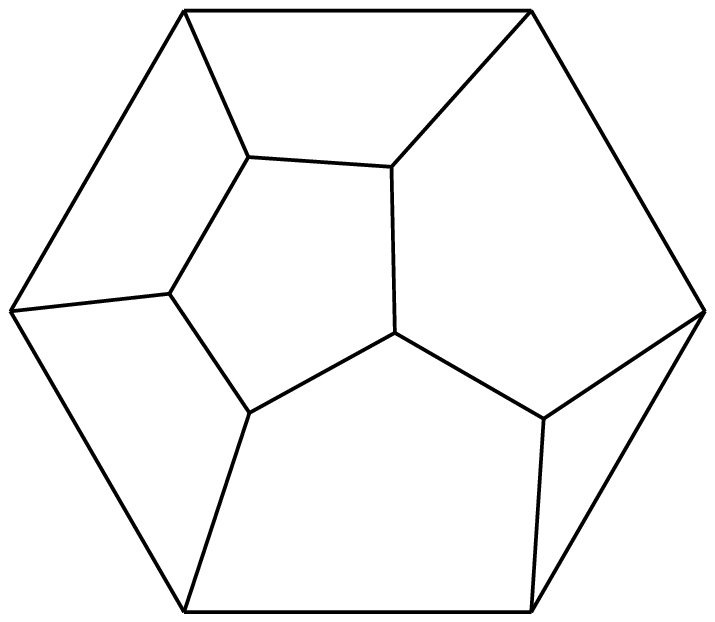}\\\hspace*{0.2mm}(00133100...)\\\hspace*{1.5mm}$F$=8, $S$=2 } & 
\specialcell[b]{P3, $f$=0.15\%\\\includegraphics[scale=0.24]{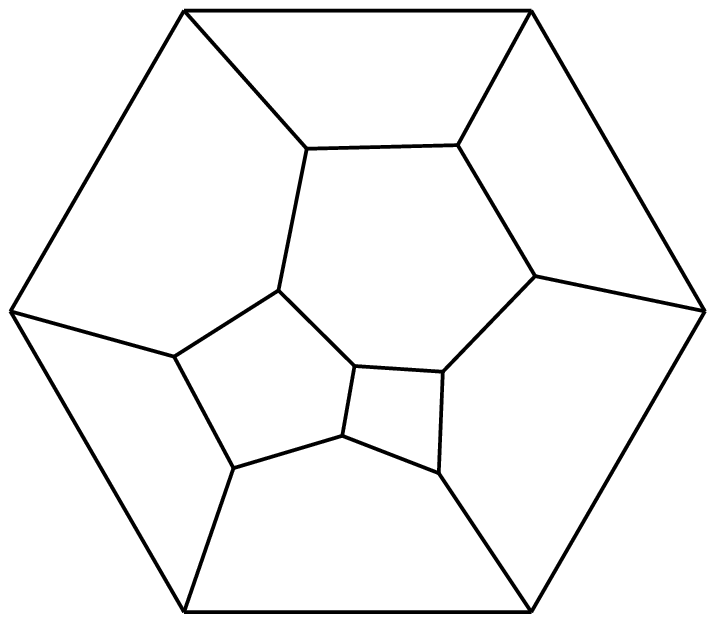}\\\hspace*{0.2mm}(00044200...)\\\hspace*{0.8mm}$F$=10, $S$=2 } & 
\specialcell[b]{P4, $f$=0.13\%\\\includegraphics[scale=0.24]{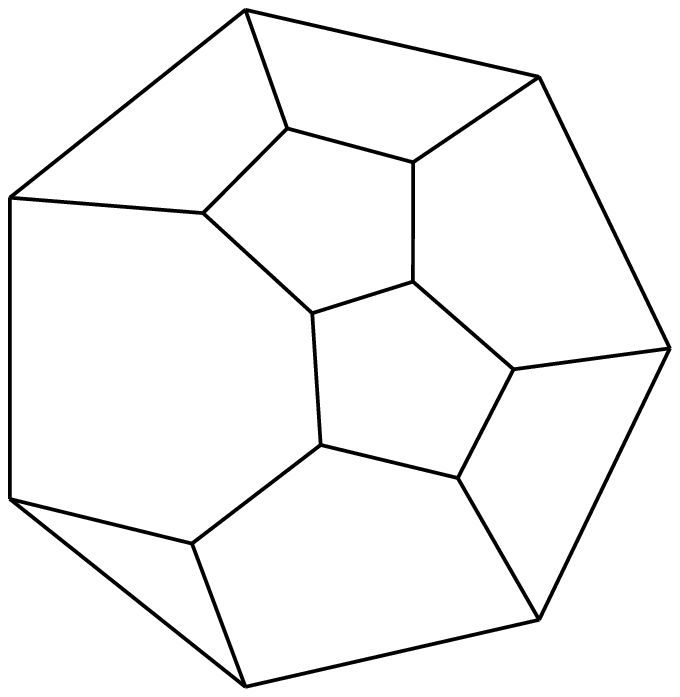}\\\hspace*{0.2mm}(00134110...)\\\hspace*{0.6mm}$F$=10, $S$=1 } & 
\specialcell[b]{P5, $f$=0.13\%\\\includegraphics[scale=0.24]{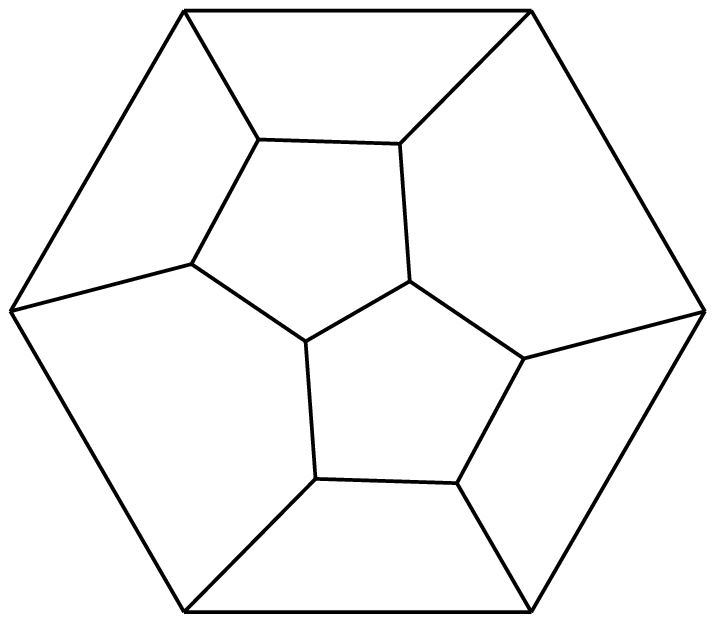}\\\hspace*{0.2mm}(00044100...)\\\hspace*{1.3mm}$F$=9, $S$=4 } & 
\specialcell[b]{P6, $f$=0.10\%\\\includegraphics[scale=0.24]{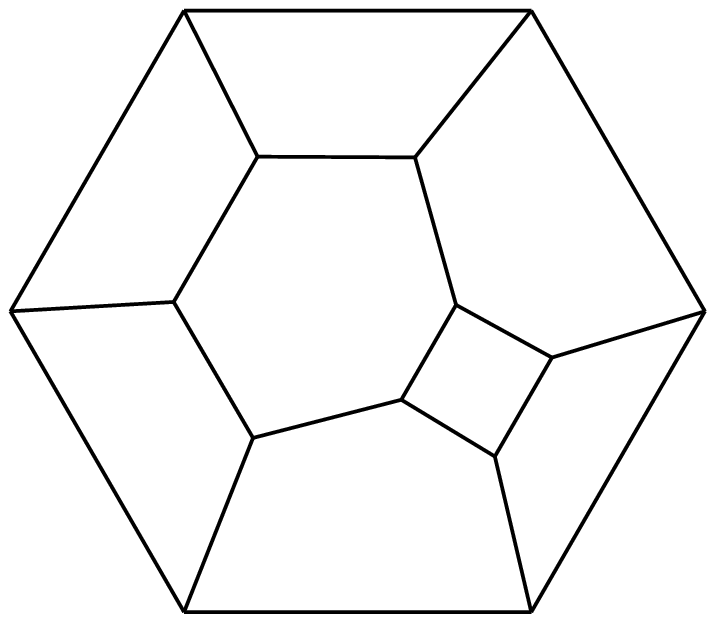}\\(00052200...)\\\hspace*{1.3mm}$F$=9, $S$=4 } & 
\specialcell[b]{P7, $f$=0.10\%\\\includegraphics[scale=0.24]{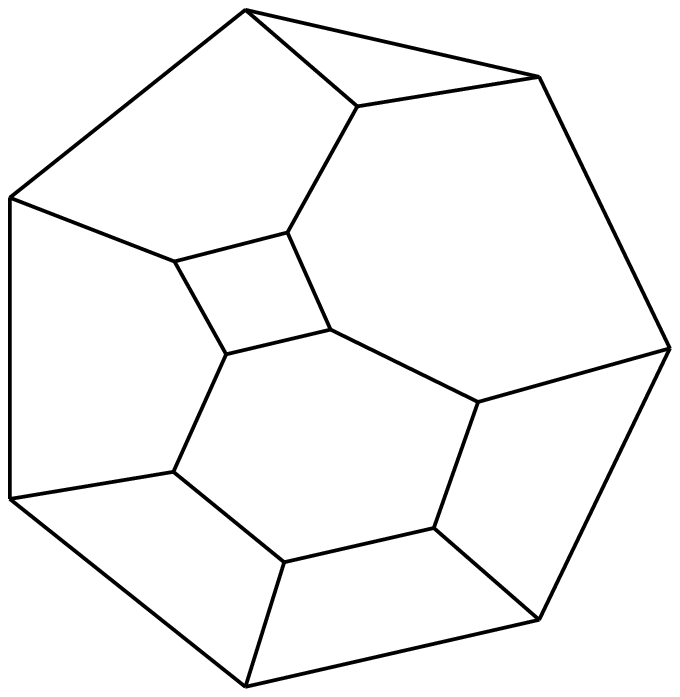}\\\hspace*{0.1mm}(00142210...)\\\hspace*{0.9mm}$F$=10, $S$=1 } & 
\specialcell[b]{P8, $f$=0.10\%\\\includegraphics[scale=0.24]{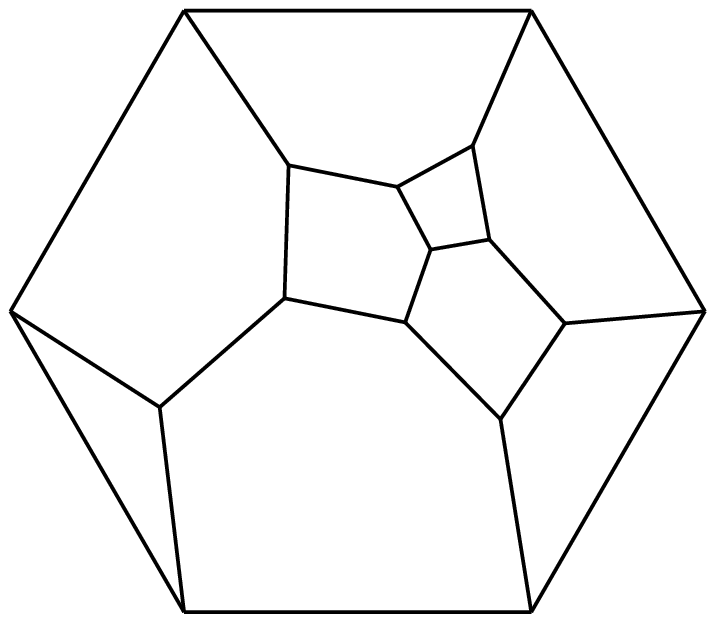}\\(00125200...)\\\hspace*{0.5mm}$F$=10, $S$=2 }\\
\hline
\hline
\multicolumn{8}{| c | }{{\bf Grain growth }}\\ 
\hline
\specialcell[t]{G1, $f$=2.83\%\\\includegraphics[scale=0.24]{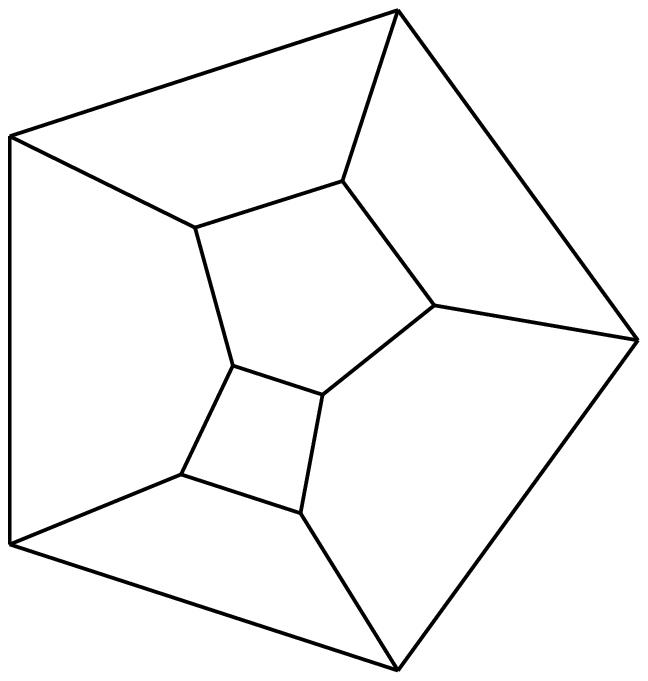}\\\hspace*{1mm}(0004400...)\\\hspace*{2mm}$F$=8, $S$=8 } & 
\specialcell[t]{G2, $f$=1.86\%\\\includegraphics[scale=0.24]{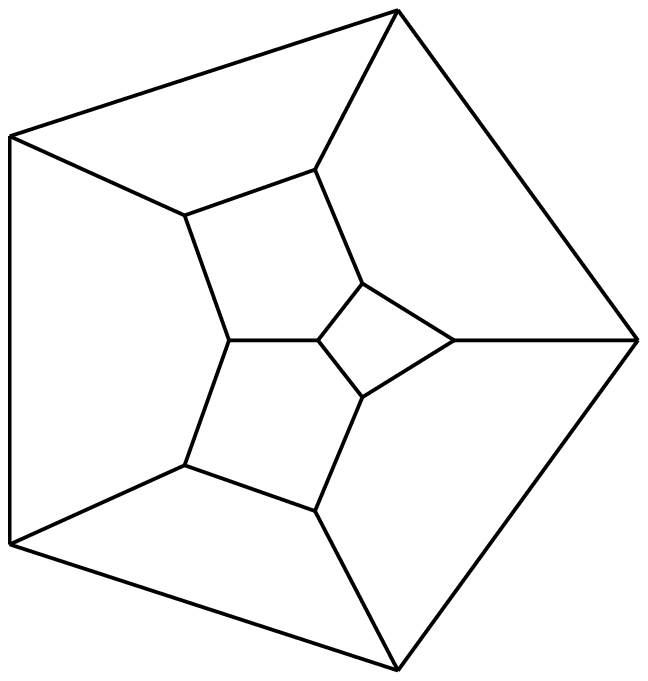}\\\hspace*{1mm}(0003600...)\\\hspace*{1mm}$F$=9, $S$=12 } & 
\specialcell[t]{G3, $f$=1.63\%\\\includegraphics[scale=0.24]{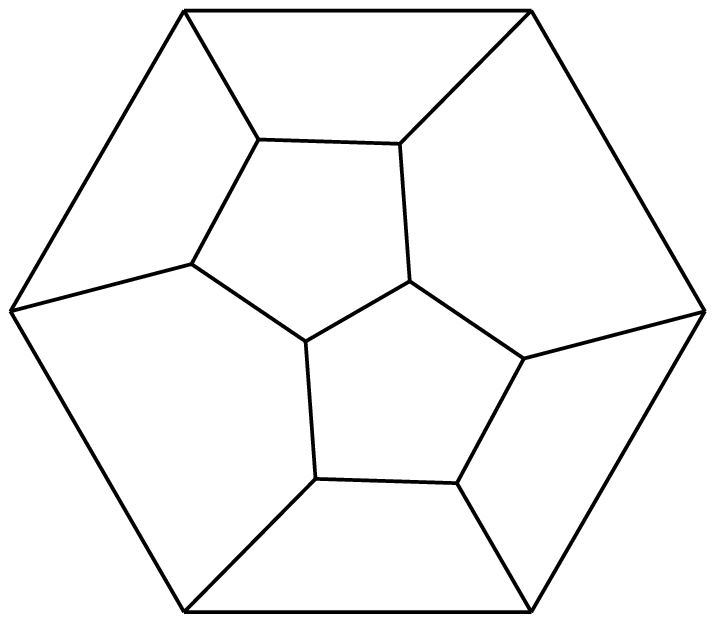}\\\hspace*{1mm}(0004410...)\\\hspace*{1.4mm}$F$=9, $S$=4 } & 
\specialcell[t]{G4, $f$=1.53\%\\\includegraphics[scale=0.24]{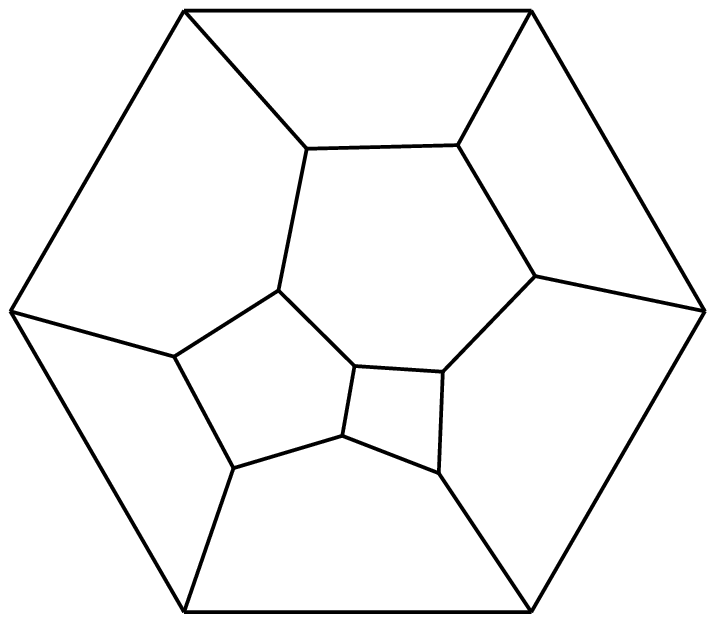}\\\hspace*{1mm}(0004420...)\\\hspace*{1mm}$F$=10, $S$=2 } & 
\specialcell[t]{G5, $f$=1.48\%\\\includegraphics[scale=0.24]{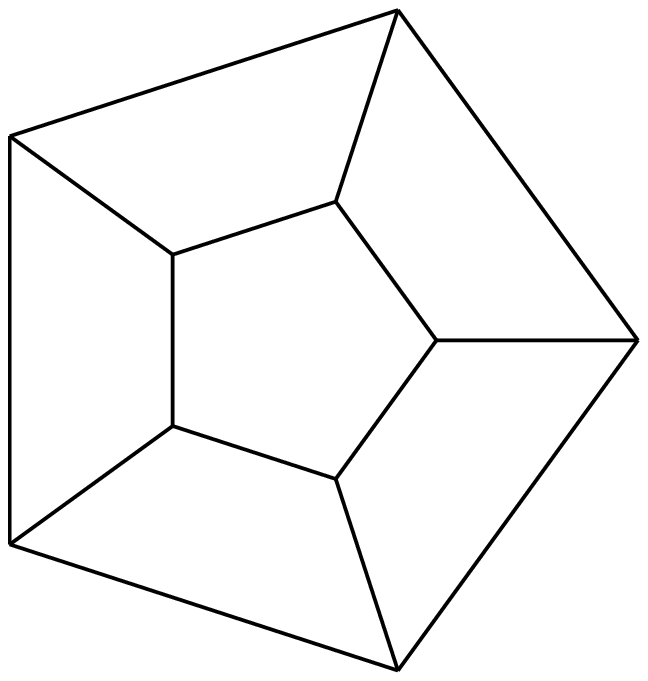}\\\hspace*{1.2mm}(0005200...)\\\hspace*{1.2mm}$F$=7, $S$=20 } & 
\specialcell[t]{G6, $f$=1.43\%\\\includegraphics[scale=0.24]{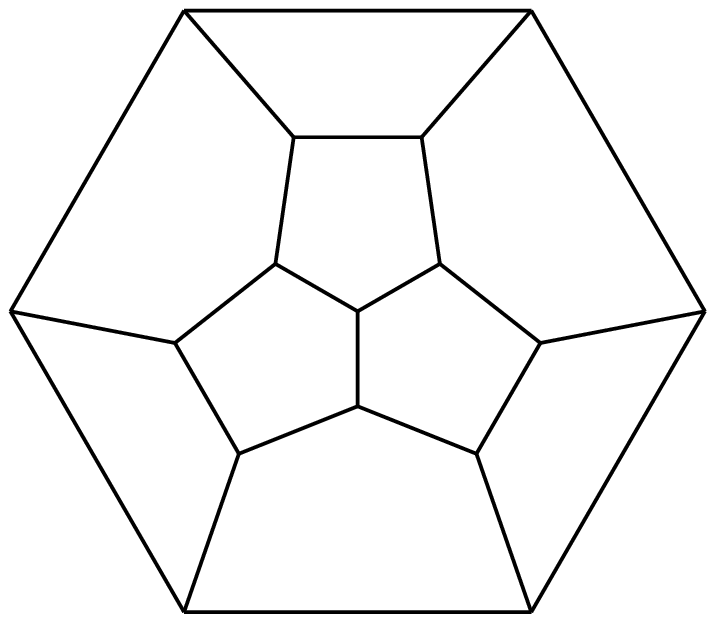}\\\hspace*{1.2mm}(0003610...)\\\hspace*{1.mm}$F$=10, $S$=6 } & 
\specialcell[t]{G7, $f$=1.39\%\\\includegraphics[scale=0.24]{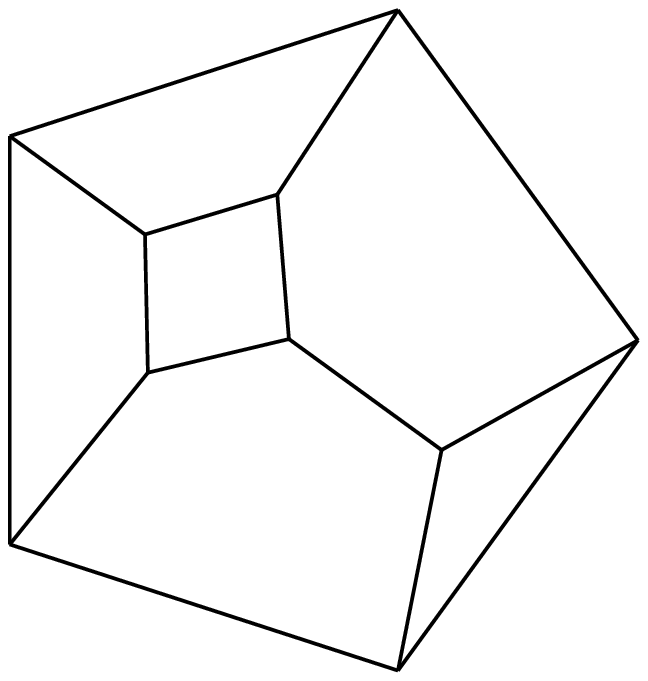}\\\hspace*{1mm}(0013300...)\\\hspace*{1.5mm}$F$=7, $S$=6 } & 
\specialcell[t]{G8, $f$=1.38\%\\\includegraphics[scale=0.24]{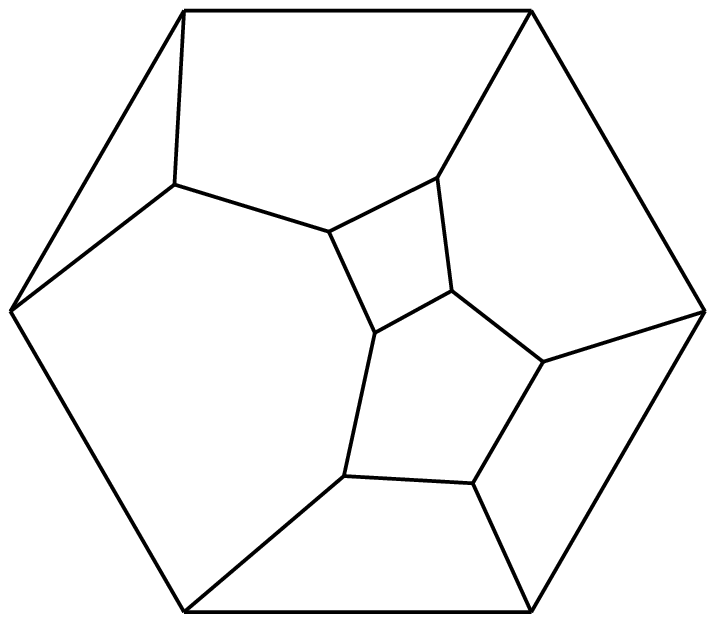}\\\hspace*{1mm}(0013320...)\\\hspace*{1.5mm}$F$=9, $S$=1 }\\
\hline
\end{tabular}
\caption{Schlegel diagrams of the 8 most common grain topologies (Weinberg vectors) in the Poisson--Voronoi and grain growth microstructures.  Listed for each topological type is a label, the frequency of occurrence $f$, the $p$-vector, the number of faces $F$, and the order $S$ of the associated symmetry group. The Weinberg vectors are tabulated in Tables SII and SIII of the Supplementary Material.}
\label{schegel-diagrams}
\end{figure*}

Schlegel diagrams associated with the eight most common Weinberg vectors in the Poisson--Voronoi and grain growth microstructures are shown in Fig.~\ref{schegel-diagrams}, along with the associated number of faces $F$, the $p$-vector, the order $S$ of the symmetry group, and their frequencies $f$; the corresponding Weinberg vectors are listed in Tables SII and SIII of the Supplementary Material.

As noted with respect to the $p$-vectors, the distribution of Weinberg vectors is much more concentrated in the grain growth microstructure than in the Poisson--Voronoi one.  The eight most common Weinberg vectors account for more than $13.5\%$ of all grains in the grain growth microstructure, but account for less than $1.35\%$ in the Poisson--Voronoi case.  Since this difference is more pronounced than for the $p$-vectors, it is clear that Weinberg vectors encode more precisely the types of grains favored by grain growth than is possible using just the number of edges around the faces of a given grain ($p$-vector information).  Indeed, a Weinberg vector specifies both the populations of these polygonal faces and their relative arrangement.  Since normal grain growth results from mean curvature flow, it drives the microstructure towards grains with smaller surface-to-volume ratios than in the Poisson--Voronoi microstructure.  Consider a soccer ball with 12 pentagonal and 20 hexagonal faces.  If the pentagonal faces were all mutually adjacent, the soccer ball would more likely be elongated and possess a larger surface-to-volume ratio.  However, grain growth drives the evolution of topology towards arrangements of faces in which pentagonal faces are separated, thus allowing for smaller surface-to-volume ratios.  While $p$-vectors do not capture this tendency, Weinberg vectors do.

To further highlight the role that Weinberg vectors play in refining $p$-vector data, consider the most common $p$-vector $(001343100...)$ in the Poisson--Voronoi microstructure, which accounts for almost $0.4\%$ the grains.  The results in Fig.~\ref{schegel-diagrams} and in Tables SII and SIII show that  none of the frequently occurring Weinberg vectors share this $p$-vector.  How is this possible?  This $p$-vector can occur in the Poisson--Voronoi microstructure in 38 topologically distinct forms!  While some of these Weinberg vectors appear over 100 times in the Poisson--Voronoi microstructure, others appear only once, if at all; such an acute disparity between the most and least frequent topological realization of this $p$-vector can also be found in grain growth microstructures.  This further illustrates that the $p$-vector alone cannot predict the frequency of a given topological type.  

\begin{figure}
\centering
\includegraphics[width=0.96\columnwidth]{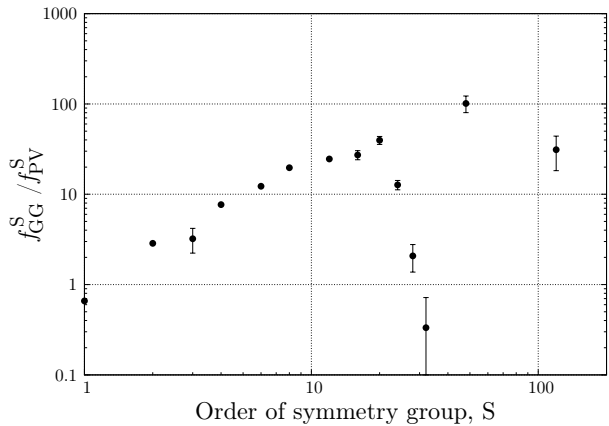}
\caption{A log-log plot of the ratio of the frequencies of grains with a given symmetry group order $S$ ($\leq 120$) from the grain growth microstructures $f^S_{GG}$ and Poisson--Voronoi microstructures $f^S_{PV}$.  Note that the statistical errors for $S = 3$, $32$, $48$, and $120$ all exceed 30\% because of their extreme rarity in the microstructures.}
\label{topology_ratios}
\end{figure} 

Figure~\ref{schegel-diagrams} and Tables SII and SIII also indicate the orders of the symmetry groups of the most frequent grain topologies.  A cursory examination reveals that the most frequent grain topologies in grain growth microstructures are substantially more symmetric than the corresponding ones for Poisson--Voronoi microstructures.  This observation is made more quantitative in Fig.~\ref{topology_ratios}.  Consider the probability of a randomly selected grain having a particular symmetry order.  The ratio of these probabilities for the grain growth and Poisson--Voronoi microstructures is plotted as a function of the order of the symmetry group in Figure \ref{topology_ratios} and summarized in Table SIV of the Supplementary Material.  These results show that complete grain topology and the frequencies of the order of  symmetries provide an outstanding tool for distinguishing between different cellular microstructures; in the present case, the differences between the relative frequencies of highly symmetric grains in the grain growth and Poisson--Voronoi microstructures can be as large as a factor of 100.  That is, highly symmetric grains are substantially more common in the grain growth microstructure. Equally interesting is that the ratio between the probability that a grain has a particular symmetry order $S$ in the grain growth and Poisson--Voronoi microstructures increases rapidly with $S$; the roughly straight curve that passes through the data points in Fig.~\ref{topology_ratios} indicates that this ratio $f^S_{GG}/f^S_{PV} \approx S^{1.2}$ (we exclude data from the fit for which the statistical error exceeds 25\% - i.e., $S = 3$, $32$, $48$, and $120$).


As with the relatively stronger selection for certain $p$-vectors and Weinberg vectors in the grain growth microstructures than in the Poisson--Voronoi microstructures, the difference in the symmetry of the grains may have its origin in the energy-minimizing process of mean curvature flow which is associated with grain growth.  While a spherical grain shape minimizes its surface area-to-volume ratio and is favored by mean curvature flow \cite{1984huisken}, grains in a cellular network must fill space, and so their faces must be polygonal.  Nevertheless, just as curvature flow drives towards geometrically symmetric spheres, we suggest that it also drives towards topologically symmetric polyhedra, as seen in the grain growth microstructures.

We have introduced an efficient method to completely classify grain topologies and have applied this method, along with $p$-vectors, to investigate some differences between Poisson--Voronoi and grain growth microstructures.  The grain growth microstructure has been observed to strongly favor certain highly symmetric grain topologies relative to the Poisson--Voronoi microstructure.  The availability of a complete topological characterization of individual cells in such cellular microstructures has proven to be an ideal tool for distinguishing between fundamental characteristics of different microstructures.  The distribution of the orders of symmetry of  grain topologies, in particular, provides a strong and convenient differentiator of different cellular structures.  

\begin{acknowledgments}
\vspace{-5mm}
EAL acknowledges support of the NSF under agreement No.~DMS-0635607.  JKM was supported under the auspices of the U.S. Department of Energy by Lawrence Livermore National Laboratory under Contract DE-AC52-07NA27344.
\end{acknowledgments}

\bibliographystyle{apsrev4-1.bst}
\bibliography{refs}       

\end{document}


\title{Supplementary Material}

\begin{abstract}
This Supplementary Material contains, in tabular form, more data describing the topology of grains from the Poisson-Voronoi and grain growth microstructures.  In particular, Table SI shows the 50 most frequent $p$-vectors in the two microstructures and their relative frequencies. Tables SII and SIII report the 50 most frequent Weinberg vectors, the order of their symmetry groups and their frequencies for the Poisson-Voronoi and grain growth microstructures, respectively.  Finally, Table SIV shows the frequencies of grains of different symmetry group orders for the two microstructures. 
\end{abstract}
\maketitle

\begin{table}
\centering
\makebox[0pt][c]{\parbox{0.8\columnwidth}{
\begin{minipage}[b]{0.45\hsize}\centering
\begin{tabular}{|c|c|c|c|}
\hline
\multicolumn{4}{|c|}{{\bf Poisson--Voronoi}} \\
\hline
& $p$-vector & $F$ & frequency \\
\hline
1 & $(001343100...)$ & 12 & 0.00391 $\pm$ 0.00012 \\
2 & $(001342100...)$ & 11 & 0.00326 $\pm$ 0.00011 \\
3 & $(001433200...)$ & 13 & 0.00295 $\pm$ 0.00010 \\
4 & $(002333110...)$ & 13 & 0.00289 $\pm$ 0.00010 \\
5 & $(001332000...)$ & 9 & 0.00283 $\pm$ 0.00010 \\
6 & $(001344100...)$ & 13 & 0.00282 $\pm$ 0.00010 \\
7 & $(001352200...)$ & 13 & 0.00282 $\pm$ 0.00010 \\
8 & $(001423100...)$ & 11 & 0.00280 $\pm$ 0.00010 \\
9 & $(002233100...)$ & 11 & 0.00270 $\pm$ 0.00010 \\
10 & $(001353200...)$ & 14 & 0.00259 $\pm$ 0.00010 \\
11 & $(001422100...)$ & 10 & 0.00257 $\pm$ 0.00010 \\
12 & $(002332110...)$ & 12 & 0.00256 $\pm$ 0.00010 \\
13 & $(000442000...)$ & 10 & 0.00255 $\pm$ 0.00010 \\
14 & $(001432200...)$ & 12 & 0.00250 $\pm$ 0.00010 \\
15 & $(000443000...)$ & 11 & 0.00243 $\pm$ 0.00009 \\
16 & $(002322200...)$ & 11 & 0.00242 $\pm$ 0.00009 \\
17 & $(001443110...)$ & 14 & 0.00238 $\pm$ 0.00009 \\
18 & $(001424100...)$ & 12 & 0.00237 $\pm$ 0.00009 \\
19 & $(002342210...)$ & 14 & 0.00236 $\pm$ 0.00009 \\
20 & $(002242200...)$ & 12 & 0.00233 $\pm$ 0.00009 \\
21 & $(001442110...)$ & 13 & 0.00230 $\pm$ 0.00009 \\
22 & $(002343210...)$ & 15 & 0.00228 $\pm$ 0.00009 \\
23 & $(000533100...)$ & 12 & 0.00216 $\pm$ 0.00009 \\
24 & $(002232100...)$ & 10 & 0.00215 $\pm$ 0.00009 \\
25 & $(002323200...)$ & 12 & 0.00215 $\pm$ 0.00009 \\
26 & $(001434200...)$ & 14 & 0.00214 $\pm$ 0.00009 \\
27 & $(002243200...)$ & 13 & 0.00213 $\pm$ 0.00009 \\
28 & $(002423210...)$ & 14 & 0.00213 $\pm$ 0.00009 \\
29 & $(001252000...)$ & 10 & 0.00210 $\pm$ 0.00009 \\
30 & $(002334110...)$ & 14 & 0.00210 $\pm$ 0.00009 \\
31 & $(002422210...)$ & 13 & 0.00206 $\pm$ 0.00009 \\
32 & $(001341100...)$ & 10 & 0.00201 $\pm$ 0.00009 \\
33 & $(001263100...)$ & 13 & 0.00200 $\pm$ 0.00009 \\
34 & $(001345100...)$ & 14 & 0.00197 $\pm$ 0.00009 \\
35 & $(001444110...)$ & 15 & 0.00195 $\pm$ 0.00008 \\
36 & $(002333300...)$ & 14 & 0.00193 $\pm$ 0.00008 \\
37 & $(001334000...)$ & 11 & 0.00193 $\pm$ 0.00008 \\
38 & $(000452100...)$ & 12 & 0.00186 $\pm$ 0.00008 \\
39 & $(001443300...)$ & 15 & 0.00183 $\pm$ 0.00008 \\
40 & $(001354200...)$ & 15 & 0.00183 $\pm$ 0.00008 \\
41 & $(002332300...)$ & 13 & 0.00179 $\pm$ 0.00008 \\
42 & $(002234100...)$ & 12 & 0.00178 $\pm$ 0.00008 \\
43 & $(001442300...)$ & 14 & 0.00178 $\pm$ 0.00008 \\
44 & $(002324200...)$ & 13 & 0.00178 $\pm$ 0.00008 \\
45 & $(000453100...)$ & 13 & 0.00175 $\pm$ 0.00008 \\
46 & $(001533210...)$ & 15 & 0.00173 $\pm$ 0.00008 \\
47 & $(001351200...)$ & 12 & 0.00173 $\pm$ 0.00008 \\
48 & $(001453210...)$ & 16 & 0.00173 $\pm$ 0.00008 \\
49 & $(002244200...)$ & 14 & 0.00171 $\pm$ 0.00008 \\
50 & $(001452210...)$ & 15 & 0.00171 $\pm$ 0.00008 \\
\hline
\end{tabular}
\end{minipage}
\hfill
\begin{minipage}[b]{0.45\hsize}\centering
\begin{tabular}{|c|c|c|c|}
\hline
\multicolumn{4}{|c|}{{\bf Grain growth}} \\
\hline
& $p$-vector & $F$ & frequency\\
\hline
1 & $(000440000...)$ & 8 & 0.02830 $\pm$ 0.00032 \\
2 & $(000442000...)$ & 10 & 0.02382 $\pm$ 0.00029 \\
3 & $(000360000...)$ & 9 & 0.01864 $\pm$ 0.00026 \\
4 & $(000443000...)$ & 11 & 0.01855 $\pm$ 0.00026 \\
5 & $(000441000...)$ & 9 & 0.01628 $\pm$ 0.00024 \\
6 & $(000520000...)$ & 7 & 0.01484 $\pm$ 0.00023 \\
7 & $(000363000...)$ & 12 & 0.01451 $\pm$ 0.00023 \\
8 & $(000361000...)$ & 10 & 0.01426 $\pm$ 0.00023 \\
9 & $(001330000...)$ & 7 & 0.01385 $\pm$ 0.00023 \\
10 & $(000362000...)$ & 11 & 0.01378 $\pm$ 0.00022 \\
11 & $(001332000...)$ & 9 & 0.01376 $\pm$ 0.00022 \\
12 & $(001331000...)$ & 8 & 0.01356 $\pm$ 0.00022 \\
13 & $(000364000...)$ & 13 & 0.01158 $\pm$ 0.00021 \\
14 & $(000281000...)$ & 11 & 0.01084 $\pm$ 0.00020 \\
15 & $(000444000...)$ & 12 & 0.00976 $\pm$ 0.00019 \\
16 & $(002220000...)$ & 6 & 0.00935 $\pm$ 0.00019 \\
17 & $(000282000...)$ & 12 & 0.00927 $\pm$ 0.00018 \\
18 & $(000600000...)$ & 6 & 0.00864 $\pm$ 0.00018 \\
19 & $(000454100...)$ & 14 & 0.00784 $\pm$ 0.00017 \\
20 & $(000280000...)$ & 10 & 0.00778 $\pm$ 0.00017 \\
21 & $(000522000...)$ & 9 & 0.00761 $\pm$ 0.00017 \\
22 & $(000453100...)$ & 13 & 0.00759 $\pm$ 0.00017 \\
23 & $(000452100...)$ & 12 & 0.00697 $\pm$ 0.00016 \\
24 & $(002300000...)$ & 5 & 0.00683 $\pm$ 0.00016 \\
25 & $(000373100...)$ & 14 & 0.00633 $\pm$ 0.00015 \\
26 & $(001252000...)$ & 10 & 0.00631 $\pm$ 0.00015 \\
27 & $(000372100...)$ & 13 & 0.00629 $\pm$ 0.00015 \\
28 & $(000533100...)$ & 12 & 0.00616 $\pm$ 0.00015 \\
29 & $(000374100...)$ & 15 & 0.00598 $\pm$ 0.00015 \\
30 & $(000451100...)$ & 11 & 0.00513 $\pm$ 0.00014 \\
31 & $(000365000...)$ & 14 & 0.00496 $\pm$ 0.00014 \\
32 & $(001343100...)$ & 12 & 0.00492 $\pm$ 0.00013 \\
33 & $(000445000...)$ & 13 & 0.00482 $\pm$ 0.00013 \\
34 & $(001342100...)$ & 11 & 0.00473 $\pm$ 0.00013 \\
35 & $(000284000...)$ & 14 & 0.00471 $\pm$ 0.00013 \\
36 & $(000455100...)$ & 15 & 0.00469 $\pm$ 0.00013 \\
37 & $(000534100...)$ & 13 & 0.00466 $\pm$ 0.00013 \\
38 & $(001341100...)$ & 10 & 0.00456 $\pm$ 0.00013 \\
39 & $(000462200...)$ & 14 & 0.00436 $\pm$ 0.00013 \\
40 & $(000283000...)$ & 13 & 0.00410 $\pm$ 0.00012 \\
41 & $(000523000...)$ & 10 & 0.00408 $\pm$ 0.00012 \\
42 & $(000375100...)$ & 16 & 0.00407 $\pm$ 0.00012 \\
43 & $(000463200...)$ & 15 & 0.00394 $\pm$ 0.00012 \\
44 & $(001422100...)$ & 10 & 0.00394 $\pm$ 0.00012 \\
45 & $(002222000...)$ & 8 & 0.00394 $\pm$ 0.00012 \\
46 & $(000464200...)$ & 16 & 0.00393 $\pm$ 0.00012 \\
47 & $(001423100...)$ & 11 & 0.00388 $\pm$ 0.00012 \\
48 & $(001333000...)$ & 10 & 0.00388 $\pm$ 0.00012 \\
49 & $(000532100...)$ & 11 & 0.00373 $\pm$ 0.00012 \\
50 & $(000371100...)$ & 12 & 0.00355 $\pm$ 0.00011 \\
\hline
\end{tabular}
\end{minipage}
}}
\caption{The 50 most frequent $p$-vectors in the Poisson--Voronoi and grain growth microstructures.  $F$ indicates the number of faces; the quoted errors indicate standard errors from the mean.}
\label{pvectortable}
\end{table}

\begin{table}
\begin{tabular}{|c|l|c|c|c|c|}
\hline
& {\bf  \text{     } Weinberg vector} & $F$ & {\bf $p$-vector} & $S$ & frequency\\
\hline
\hline
\multicolumn{6}{|c|}{{\bf Poisson--Voronoi}} \\
\hline
1 & ABCACDEFAFGHIBIJKDKLELMGMNHNJNMLKJIHGFEDCB & 9 & (0133200...)  & 1 & $ 0.00283 \pm 0.00010 $ \\
2 & ABCACDEFAFGHBHIJDJKEKLGLILKJIHGFEDCB & 8 & (0133100...)  & 2 & $ 0.00170 \pm 0.00008 $ \\
3 & ABCDADEFAFGHIBIJKCKLMEMNGNOHOPJPLPONMLKJIHGFEDCB & 10 & (0044200...)  & 2 & $ 0.00150 \pm 0.00007 $ \\
4 & ABCACDEFAFGHIBIJKLDLMEMNGNOPHPJPOKONMLKJIHGFEDCB & 10 & (0134110...)  & 1 & $ 0.00131 \pm 0.00007 $ \\
5 & ABCDADEFAFGHBHIJCJKLELMGMNINKNMLKJIHGFEDCB & 9 & (0044100...)  & 4 & $ 0.00129 \pm 0.00007 $ \\
6 & ABCDADEFAFGHBHIJKCKLELMNGNINMJMLKJIHGFEDCB & 9 & (0052200...)  & 4 & $ 0.00101 \pm 0.00006 $ \\
7 & ABCACDEFAFGHIBIJKLDLMNENGNMOPHPJPOKOMLKJIHGFEDCB & 10 & (0142210...)  & 1 & $ 0.00099 \pm 0.00006 $ \\
8 & ABCACDEFAFGHIBIJKDKLMEMNGNOPHPJPOLONMLKJIHGFEDCB & 10 & (0125200...)  & 2 & $ 0.00095 \pm 0.00006 $ \\
9 & ABCDADEFAFGHBHIJCJKEKLGLILKJIHGFEDCB & 8 & (0044000...)  & 8 & $ 0.00095 \pm 0.00006 $ \\
10 & ABCACDEFAFGHIBIJKLDLMEMNOGOPHPJPONKNMLKJIHGFEDCB & 10 & (0142210...)  & 1 & $ 0.00095 \pm 0.00006 $ \\
11 & ABCDADEFAFGHBHIJKCKLMEMNOGOPIPQJQRLRNRQPONMLKJIH& 11 & (0044300...)  & 1 & $ 0.00092 \pm 0.00006 $ \\
12 & ABCACDEFAFGHIBIJKDKLELMNGNOHOPJPMPONMLKJIHGFEDCB & 10 & (0133300...)  & 1 & $ 0.00090 \pm 0.00006 $ \\
13 & ABCACDEFAFGHBHIJDJKLELMGMNINKNMLKJIHGFEDCB & 9 & (0125100...)  & 2 & $ 0.00089 \pm 0.00006 $ \\
14 & ABCDADEFAFGHBHIJCJKLELMNGNOIOPKPMPONMLKJIHGFEDCB & 10 & (0044200...)  & 2 & $ 0.00088 \pm 0.00006 $ \\
15 & ABCACDEFGAGHIJBJKLMDMNENOPFPHPOQRIRKRQLQONMLKJIH& 11 & (0142310...)  & 1 & $ 0.00088 \pm 0.00006 $ \\
16 & ABCACDEFAFGHIBIJKLDLMNENOGOPKPMPONMLKJHJIHGFEDCB & 10 & (0223210...)  & 1 & $ 0.00086 \pm 0.00006 $ \\
17 & ABCDADEFGAGHIBIJKCKLMEMNFNOPHPQJQRLRORQPONMLKJIH& 11 & (0036200...)  & 2 & $ 0.00085 \pm 0.00006 $ \\
18 & ABCACDEFAFGHIBIJKLDLMNENOGOPQHQJQPRKRMRPONMLKJIH& 11 & (0126110...)  & 1 & $ 0.00085 \pm 0.00006 $ \\
19 & ABCACDEFAFGHIBIJKLDLMEMNGNKNMLKJHJIHGFEDCB & 9 & (0223110...)  & 2 & $ 0.00079 \pm 0.00005 $ \\
20 & ABCACDEFAFGHIBIJKDKLMEMNGNOHOPJPLPONMLKJIHGFEDCB & 10 & (0125200...)  & 2 & $ 0.00079 \pm 0.00005 $ \\
21 & ABCDADEFAFGHIBIJKCKLMEMNOGOPHPQJQRLRNRQPONMLKJIH& 11 & (0044300...)  & 2 & $ 0.00078 \pm 0.00005 $ \\
22 & ABCACDEFAFGHIJBJKLMDMNENOPGPHPOLONMLKIKJIHGFEDCB & 10 & (0321220...)  & 1 & $ 0.00076 \pm 0.00005 $ \\
23 & ABCACDEFGAGHIJBJKLDLMEMNFNOHOPIPKPONMLKJIHGFEDCB & 10 & (0141400...)  & 2 & $ 0.00074 \pm 0.00005 $ \\
24 & ABCACDEFAFGHIBIJKLDLMNENGNMKMLKJHJIHGFEDCB & 9 & (0312210...)  & 2 & $ 0.00074 \pm 0.00005 $ \\
25 & ABCDADEFGAGHIBIJKCKLMEMNFNOHOPJPLPONMLKJIHGFEDCB & 10 & (0036100...)  & 6 & $ 0.00072 \pm 0.00005 $ \\
26 & ABCACDEFAFGHIBIJKLDLMEMNGNOHOPJPKPONMLKJIHGFEDCB & 10 & (0223210...)  & 1 & $ 0.00071 \pm 0.00005 $ \\
27 & ABCDADEFGAGHIBIJKCKLELMFMNHNJNMLKJIHGFEDCB & 9 & (0036000...)  & 12 & $ 0.00067 \pm 0.00005 $ \\
28 & ABCDADEFAFGHBHIJCJKLMEMNGNOPIPKPOLONMLKJIHGFEDCB & 10 & (0053110...)  & 2 & $ 0.00066 \pm 0.00005 $ \\
29 & ABCDADEFGAGHIBIJKCKLMEMNOFOPQHQRJRSTLTNTSPSRQPON& 12 & (0036300...)  & 1 & $ 0.00064 \pm 0.00005 $ \\
30 & ABCACDEFAFGHIBIJKDKLMEMNOGOPHPQJQRLRNRQPONMLKJIH& 11 & (0125300...)  & 1 & $ 0.00064 \pm 0.00005 $ \\
31 & ABCACDEFAFGHIBIJKDKLELGLKJHJIHGFEDCB & 8 & (0222200...)  & 4 & $ 0.00064 \pm 0.00005 $ \\
32 & ABCACDEFAFGHBHIJKDKLELMNGNINMJMLKJIHGFEDCB & 9 & (0142110...)  & 2 & $ 0.00064 \pm 0.00005 $ \\
33 & ABCDADEFGAGHIBIJKCKLMEMNOFOPHPQJQRLRNRQPONMLKJIH& 11 & (0028100...)  & 4 & $ 0.00063 \pm 0.00005 $ \\
34 & ABCACDEFAFGHBHIDIJEJGJIHGFEDCB & 7 & (0133000...)  & 6 & $ 0.00062 \pm 0.00005 $ \\
35 & ABCACDEFGAGHIJBJKLMDMEMLNOFOHONPIPKPNLKJIHGFEDCB & 10 & (0231310...)  & 2 & $ 0.00061 \pm 0.00005 $ \\
36 & ABCDADEFAFGHBHIJCJKLMEMNOGOPQIQKQPRLRNRPONMLKJIH& 11 & (0053210...)  & 1 & $ 0.00059 \pm 0.00005 $ \\
37 & ABCACDEFAFGHIBIJKLDLMNENOGOPQRHRJRQKQPMPONMLKJIH& 11 & (0134210...)  & 1 & $ 0.00059 \pm 0.00005 $ \\
38 & ABCDADEFAFGHIBIJKCKLMNENOGOPQHQRJRLRQPMPONMLKJIH& 11 & (0045110...)  & 2 & $ 0.00058 \pm 0.00005 $ \\
39 & ABCACDEFGAGHIJBJKLDLMEMNOFOPHPQIQRKRNRQPONMLKJIH& 11 & (0125300...)  & 2 & $ 0.00055 \pm 0.00005 $ \\
40 & ABCACDEFGAGHIJBJKLMDMEMLNOFOPHPQIQRKRNRQPONLKJIH& 11 & (0223310...)  & 1 & $ 0.00055 \pm 0.00005 $ \\
41 & ABCACDEFAFGHBHIJKDKLELMNGNOIOPJPMPONMLKJIHGFEDCB & 10 & (0134110...)  & 1 & $ 0.00054 \pm 0.00004 $ \\
42 & ABCACDEAEFGHBHIJKDKLFLJLKJIGIHGFEDCB & 8 & (0312110...)  & 1 & $ 0.00053 \pm 0.00004 $ \\
43 & ABCACDEFAFGHIJBJKLMDMNENOGOPHPLPONMLKIKJIHGFEDCB & 10 & (0224020...)  & 2 & $ 0.00053 \pm 0.00004 $ \\
44 & ABCACDEFAFGHIBIJKLDLMEMNOGOPHPQJQRKRNRQPONMLKJIH& 11 & (0134210...)  & 1 & $ 0.00053 \pm 0.00004 $ \\
45 & ABCACDEFAFGHIBIJKDKLMEMNOGOPQHQJQPRLRNRPONMLKJIH& 11 & (0133400...)  & 1 & $ 0.00053 \pm 0.00004 $ \\
46 & ABCACDEAEFGHBHIJKDKLFLMNGNINMJMLKJIHGFEDCB & 9 & (0142110...)  & 1 & $ 0.00053 \pm 0.00004 $ \\
47 & ABCDADEFAFGHIBIJKCKLMNENOPGPQHQRSJSLSRTMTOTRQPON& 12 & (0053310...)  & 1 & $ 0.00052 \pm 0.00004 $ \\
48 & ABCACDEFGAGHIJBJKLDLELKMNFNHNMIMKJIHGFEDCB & 9 & (0230400...)  & 4 & $ 0.00051 \pm 0.00004 $ \\
49 & ABCDADEFAFGHBHIJKCKLMEMNOGOIONPJPLPNMLKJIHGFEDCB & 10 & (0052300...)  & 4 & $ 0.00051 \pm 0.00004 $ \\
50 & ABCACDEFGAGHIJBJKLMDMNENOFOPHPQRIRKRQLQPONMLKJIH& 11 & (0142310...)  & 1 & $ 0.00051 \pm 0.00004 $ \\
\hline
\end{tabular}
\caption{The 50 most common Weinberg vectors and their number of faces $F$,  $p$-vectors, the order $S$ of the associated symmetry group, and frequencies in the Poisson--Voronoi microstructure.  To consolidate the data, integers are replaced with letters (e.g., 1 with A, 2 with B, etc.).  Long Weinberg vectors were  truncated;  in these cases, the remainder of the characters are the first characters of the alphabet, in reverse order, up to the last letter shown.   The quoted errors indicate standard errors from the mean.}
\label{wvectortable-pv}
\end{table}

\begin{table}
\begin{tabular}{|c|l|c|l|c|c|}
\hline
& {\bf  \text{     } Weinberg vector} & $F$ & {\bf $p$-vector} & $S$ & frequency\\
\hline
\hline
\multicolumn{6}{|c|}{{\bf Grain growth}} \\
\hline
1 & ABCDADEFAFGHBHIJCJKEKLGLILKJIHGFEDCB & 8 & (0044000) & 8 & $ 0.02830 \pm 0.00032 $ \\
2 & ABCDADEFGAGHIBIJKCKLELMFMNHNJNMLKJIHGFEDCB & 9 & (0036000) & 12 & $ 0.01864 \pm 0.00026 $ \\
3 & ABCDADEFAFGHBHIJCJKLELMGMNINKNMLKJIHGFEDCB & 9 & (0044100) & 4 & $ 0.01628 \pm 0.00024 $ \\
4 & ABCDADEFAFGHIBIJKCKLMEMNGNOHOPJPLPONMLKJIHGFEDCB & 10 & (0044200) & 2 & $ 0.01525 \pm 0.00024 $ \\
5 & ABCDADEFAFGHBHICIJEJGJIHGFEDCB & 7 & (0052000) & 20 & $ 0.01484 \pm 0.00023 $ \\
6 & ABCDADEFGAGHIBIJKCKLMEMNFNOHOPJPLPONMLKJIHGFEDCB & 10 & (0036100) & 6 & $ 0.01426 \pm 0.00023 $ \\
7 & ABCACDEFAFGHBHIDIJEJGJIHGFEDCB & 7 & (0133000) & 6 & $ 0.01385 \pm 0.00023 $ \\
8 & ABCACDEFAFGHIBIJKDKLELMGMNHNJNMLKJIHGFEDCB & 9 & (0133200) & 1 & $ 0.01376 \pm 0.00022 $ \\
9 & ABCACDEFAFGHBHIJDJKEKLGLILKJIHGFEDCB & 8 & (0133100) & 2 & $ 0.01356 \pm 0.00022 $ \\
10 & ABCDADEFGAGHIBIJKCKLMEMNFNOPHPQJQRLRORQPONMLKJIH & 11 & (0036200) & 2 & $ 0.01144 \pm 0.00020 $ \\
11 & ABCDADEFGAGHIBIJKCKLMEMNOFOPHPQJQRLRNRQPONMLKJIH & 11 & (0028100) & 4 & $ 0.01084 \pm 0.00020 $ \\
12 & ABCACDEAEFGBGHDHFHGFEDCB & 6 & (0222000) & 4 & $ 0.00935 \pm 0.00019 $ \\
13 & ABCDADEFAFGBGHCHEHGFEDCB & 6 & (0060000) & 48 & $ 0.00864 \pm 0.00018 $ \\
14 & ABCDADEFGAGHIBIJKCKLELMNFNOHOPJPMPONMLKJIHGFEDCB & 10 & (0028000) & 16 & $ 0.00778 \pm 0.00017 $ \\
15 & ABCDADEFAFGHBHIJCJKLELMNGNOIOPKPMPONMLKJIHGFEDCB & 10 & (0044200) & 2 & $ 0.00766 \pm 0.00017 $ \\
16 & ABCDADEFAFGHBHIJKCKLELMNGNINMJMLKJIHGFEDCB & 9 & (0052200) & 4 & $ 0.00761 \pm 0.00017 $ \\
17 & ABCACDEAEFBFDFEDCB & 5 & (0230000) & 12 & $ 0.00683 \pm 0.00016 $ \\
18 & ABCDADEFGAGHIBIJKCKLMEMNOFOPQHQRJRSTLTNTSPSRQPON& 12 & (0036300) & 1 & $ 0.00650 \pm 0.00015 $ \\
19 & ABCDADEFAFGHBHIJKCKLMEMNOGOPIPQJQRLRNRQPONMLKJIH & 11 & (0044300) & 1 & $ 0.00642 \pm 0.00015 $ \\
20 & ABCDADEFAFGHIBIJKCKLMEMNOGOPHPQJQRLRNRQPONMLKJIH & 11 & (0044300) & 2 & $ 0.00637 \pm 0.00015 $ \\
21 & ABCDADEFGAGHIBIJKLCLMNENOFOPQHQJQPRKRMRPONMLKJIH & 11 & (0044300) & 4 & $ 0.00498 \pm 0.00014 $ \\
22 & ABCDADEFGAGHIBIJKCKLMEMNOFOPHPQRJRSLSTNTQTSRQPON& 12 & (0028200) & 2 & $ 0.00480 \pm 0.00013 $ \\
23 & ABCDADEFGAGHIBIJKLCLMNENOPFPQHQRJRSKSTMTOTSRQPON& 12 & (0036300) & 2 & $ 0.00465 \pm 0.00013 $ \\
24 & ABCDADEFGAGHIBIJKLCLMNENOPFPQHQRJRSTKTMTSOSRQPON& 12 & (0028200) & 4 & $ 0.00446 \pm 0.00013 $ \\
25 & ABCDADEFAFGHBHIJKCKLMEMNOGOIONPJPLPNMLKJIHGFEDCB & 10 & (0052300) & 4 & $ 0.00408 \pm 0.00012 $ \\
26 & ABCACDEFAFGHIBIJKDKLELGLKJHJIHGFEDCB & 8 & (0222200) & 4 & $ 0.00390 \pm 0.00012 $ \\
27 & ABCACDEFAFGHIBIJKLDLMEMNGNOPHPJPOKONMLKJIHGFEDCB & 10 & (0134110) & 1 & $ 0.00368 \pm 0.00012 $ \\
28 & ABCDADEFGAGHIBIJKLCLMNENOPFPQRHRSJSTUKUMUTVOVQVT& 13 & (0036400) & 1 & $ 0.00347 \pm 0.00011 $ \\
29 & ABCACDEFAFGHIBIJKDKLMEMNGNOPHPJPOLONMLKJIHGFEDCB & 10 & (0125200) & 2 & $ 0.00331 \pm 0.00011 $ \\
30 & ABCDADEFGAGHIBIJKLCLMNENOFOPQHQRJRSTKTMTSPSRQPON& 12 & (0044400) & 2 & $ 0.00328 \pm 0.00011 $ \\
31 & ABCDADEFAFGHIBIJKCKLMNENOGOPQHQRJRLRQPMPONMLKJIH & 11 & (0045110) & 2 & $ 0.00327 \pm 0.00011 $ \\
32 & ABCACDEFAFGHBHIJDJKLELMGMNINKNMLKJIHGFEDCB & 9 & (0125100) & 2 & $ 0.00317 \pm 0.00011 $ \\
33 & ABCACDEAEFGHBHIJDJKFKLGLILKJIHGFEDCB & 8 & (0141200) & 2 & $ 0.00301 \pm 0.00011 $ \\
34 & ABCDADEFGAGHIJBJKLCLMNENOPFPQHQRSISTKTUMUVOVRVUT& 13 & (001\underline{10}200) & 4 & $ 0.00266 \pm 0.00010 $ \\
35 & ABCACDEFAFGHIBIJKDKLMEMNGNOHOPJPLPONMLKJIHGFEDCB & 10 & (0125200) & 2 & $ 0.00259 \pm 0.00010 $ \\
36 & ABCDADEFAFGHIBIJKCKLMEMNGNOPHPQJQRLRORQPONMLKJIH & 11 & (0036200) & 4 & $ 0.00234 \pm 0.00009 $ \\
37 & ABCACDEFAFGHIBIJKLDLMNENOGOPQHQJQPRKRMRPONMLKJIH & 11 & (0126110) & 1 & $ 0.00231 \pm 0.00009 $ \\
38 & ABCDADEFGAGHIBIJKLCLMNENOPFPQHQRSJSTUKUMUTVOVRVT& 13 & (0036400) & 1 & $ 0.00230 \pm 0.00009 $ \\
39 & ABCDADEFAFGHIBIJKLCLMNENOPGPQHQRJRSTKTMTSOSRQPON& 12 & (0044400) & 2 & $ 0.00229 \pm 0.00009 $ \\
40 & ABCACDEFAFGHIBIJKLDLMEMNGNKNMLKJHJIHGFEDCB & 9 & (0223110) & 2 & $ 0.00226 \pm 0.00009 $ \\
41 & ABCDADEFAFGHIBIJKCKLMEMNOGOPHPQRJRSLSTNTQTSRQPON& 12 & (0044400) & 1 & $ 0.00219 \pm 0.00009 $ \\
42 & ABCDADEFGAGHIBIJKCKLMNENOFOPQHQRJRSLSTMTPTSRQPON& 12 & (0037110) & 2 & $ 0.00216 \pm 0.00009 $ \\
43 & ABCDADEFAFGHBHIJCJKLMEMNGNOPIPKPOLONMLKJIHGFEDCB & 10 & (0053110) & 2 & $ 0.00214 \pm 0.00009 $ \\
44 & ABCDADEFGAGHIBIJKLCLMNOEOPFPQRHRSJSTUKUMUTVNVQVT& 13 & (0037210) & 1 & $ 0.00213 \pm 0.00009 $ \\
45 & ABCACDEAEFGHBHIJDJFJIGIHGFEDCB & 7 & (0230200) & 4 & $ 0.00209 \pm 0.00009 $ \\
46 & ABCDADEFAFGHIBIJKCKLMEMNOGOPHPQRJRLRQNQPONMLKJIH & 11 & (0052400) & 2 & $ 0.00203 \pm 0.00009 $ \\
47 & ABCACDEFAFGHBHIJDJEJIGIHGFEDCB & 7 & (0303100) & 6 & $ 0.00202 \pm 0.00009 $ \\
48 & ABCDADEFAFGHIBIJKCKLMNENOPGPQHQRSJSLSRTMTOTRQPON& 12 & (0053310) & 1 & $ 0.00202 \pm 0.00009 $ \\
49 & ABCACDEFGAGHIJBJKLDLMEMNOFOHONPIPKPNMLKJIHGFEDCB & 10 & (0133300) & 6 & $ 0.00200 \pm 0.00009 $ \\
50 & ABCACDEAEFGBGHIDIJFJHJIHGFEDCB & 7 & (0222100) & 2 & $ 0.00193 \pm 0.00008 $ \\
\hline
\end{tabular}
\caption{The 50 most common Weinberg vectors and their number of faces $F$,  $p$-vectors, the symmetry group order $S$, and frequencies in the grain growth microstructure.  To consolidate the data, integers are replaced with letters (e.g., 1 with A, 2 with B, etc.).  Long Weinberg vectors were  truncated;  in these cases, the remainder of the characters are the first characters of the alphabet, in reverse order, up to the last letter shown.  On row 34, the characters \underline{10} indicates a 10.  The quoted errors indicate standard errors from the mean.}
\label{wvectortable-gg}
\end{table}

\begin{table}
\begin{tabular}{|r|l|c|r|r|r|}
\hline
S & $f^S_{PV}$ & $f^S_{GG}$ & $f^S_{GG}/f^S_{PV}$ & $\sigma$\hspace*{3mm} & RSE\hspace*{2mm} \\
\hline
1 & 0.917583 & 0.605301 & 0.660 & 0.001 & 0.166 \\ 
2 & 0.065560 & 0.187631 & 2.862 & 0.024 & 0.830 \\ 
3 & 0.000052 & 0.000167 & 3.214 & 0.984 & 30.601 \\ 
4 & 0.010157 & 0.077951 & 7.674 & 0.155 & 2.013 \\ 
6 & 0.002812 & 0.034460 & 12.255 & 0.462 & 3.768 \\ 
8 & 0.001580 & 0.031111 & 19.685 & 0.976 & 4.959 \\ 
10 & 0.000019 & 0.000000 & 0.000 & n/a & n/a \\ 
12 & 0.001124 & 0.027672 & 24.617 & 1.441 & 5.854 \\ 
16 & 0.000286 & 0.007783 & 27.247 & 3.161 & 11.600 \\ 
20 & 0.000375 & 0.014850 & 39.634 & 3.992 & 10.071 \\ 
24 & 0.000286 & 0.003632 & 12.714 & 1.504 & 11.833 \\ 
28 & 0.000048 & 0.000100 & 2.077 & 0.701 & 33.757 \\ 
32 & 0.000011 & 0.000004 & 0.333 & 0.385 & 115.470 \\ 
48 & 0.000085 & 0.008644 & 101.304 & 21.226 & 20.952 \\ 
120 & 0.000022 & 0.000694 & 31.167 & 12.926 & 41.474 \\
\hline
\end{tabular}
\caption{Frequencies of grains with various symmetry group orders S in the Poisson--Voronoi and grain growth microstructures.   The estimated error in these ratios is given by $\sigma$; RSE indicates the relative standard error, expressed as a percentage.
}
\label{wvectortable}
\end{table}